\documentclass[lettersize,journal]{IEEEtran}
\usepackage{amsmath,amsfonts}
\usepackage{algorithmic}
\usepackage{algorithm}
\usepackage{array}
\usepackage[caption=false, font=normalsize,labelfont=sf, textfont=sf]{subfig}
\usepackage{textcomp}
\usepackage{stfloats}
\usepackage{url}
\usepackage{verbatim}
\usepackage{graphicx}
\usepackage{cite}
\usepackage{xcolor}
\usepackage[hidelinks]{hyperref}
\begin{document}

\title{ UAV Object Detection and Positioning in a Mining Industrial Metaverse with Custom Geo-Referenced Data}%UAV-Based Geo-Referenced Data Collection for Enhanced Object Detection and Positioning in a Mining Industrial Metaverse}

\author{Vasiliki Balaska*, Ioannis Tsampikos Papapetros, Katerina Maria Oikonomou, Loukas Bampis, and Antonios Gasteratos%
\thanks{ V. Balaska(*corresponding author), I.T Papapetros, K.M Oikonomou and A. Gasteratos are with the Department of Production and Management Engineering, Democritus University of Thrace, Xanthi, Greece. 
        {\tt\small vbalaska@pme.duth.gr, ipapapet@pme.duth.gr, aioikono@pme.duth.gr, agaster@pme.duth.gr}}%
        
\thanks{L. Bampis is with the Department of Electrical and Computer Engineering, Democritus University of Thrace, Xanthi, Greece. 
        {\tt\small lbampis@ee.duth.gr}}}

\markboth{Journal of \LaTeX\ Class Files,~Vol.~14, No.~8, August~2021}%
{Shell \MakeLowercase{\textit{et al.}}: A Sample Article Using IEEEtran.cls for IEEE Journals}


\maketitle

\begin{abstract}
The mining sector increasingly adopts digital tools to improve operational efficiency, safety, and data-driven decision-making. One of the key challenges remains the reliable acquisition of high-resolution, geo-referenced spatial information to support core activities such as extraction planning and on-site monitoring. This work presents an integrated system architecture that combines UAV-based sensing, LiDAR terrain modeling, and deep learning-based object detection to generate spatially accurate information for open-pit mining environments. The proposed pipeline includes geo-referencing, 3D reconstruction, and object localization, enabling structured spatial outputs to be integrated into an industrial digital twin platform. Unlike traditional static surveying methods, the system offers higher coverage and automation potential, with modular components suitable for deployment in real-world industrial contexts. While the current implementation operates in post-flight batch mode, it lays the foundation for real-time extensions. The system contributes to the development of AI-enhanced remote sensing in mining by demonstrating a scalable and field-validated geospatial data workflow that supports situational awareness and infrastructure safety.

\end{abstract}


\section{Introduction}
\IEEEPARstart{T}{he} mining industry is significantly transforming by integrating emerging digital technologies. One of the primary challenges facing this sector is the lack of high-precision real-time geospatial data to support decision-making in exploration, extraction, and safety monitoring~\cite{stothard2023application,flores2024technological}. Traditional data collection methods often involve high costs, time-consuming processes, and potential safety risks. As part of the MASTERMINE Horizon Europe project (GA 101091895)\footnote{ https://www.mastermine-project.eu/} initiative, this study focuses on creating a detailed 3D model of the mining site by integrating LiDAR scanning and UAV-based imaging. The proposed approach enables the detection of key objects using onboard cameras and deep learning techniques, followed by their projection onto the 3D map for enhanced situational awareness. Additionally, the system leverages geo-referenced images to support visual navigation, improving UAV positioning within the mining environment. By combining these elements, our work contributes to a more accurate and efficient mapping process, supporting safer and more informed decision-making in mining operations.%To address these challenges, through the Horizon Europe MASTERMINE project (GA 101091895), we leverage unmanned aerial vehicles (UAVs), geo-referenced data augmentation and integration of 3d map-model into Mining Digital Twin to create a comprehensive solution for the mining industry.

A key aspect of modern mining operations is the ability to process and analyze heterogeneous datasets collected from various sources, including UAV-mounted LiDAR, RGB imaging, Internet of Things (IoT) enabled ground sensors, and other components. However, efficiently integrating and utilizing these datasets remains a complex issue due to limitations in data interoperability, computational processing, and real-time visualization ~\cite{ghahramanieisalou2024digital}. The MASTERMINE project proposes an innovative approach that augments UAV-collected data through geospatial referencing and real-time digital modeling, enabling the creation of a Metaverse-driven virtual mining environment. This environment aims to bridge the gap between physical mining operations and digital simulations, providing an interactive and data-driven framework for decision-making.

The deployment of UAVs, robotics, and digital twins in the mining industry presents several technical and operational challenges. One of the most pressing issues is ensuring data accuracy and reliability, as UAVs and robotic systems capture extensive geospatial datasets that require high precision in geo-referencing for meaningful interpretation. Furthermore, the vast amount of data collected from aerial and ground-based sensors necessitates advanced AI-driven data processing, storage, and visualization techniques to extract actionable insights \cite{thangiartificial}. AI techniques play a pivotal role in digital transformation by enabling automated data fusion, anomaly detection, and predictive modeling, thereby enhancing operational efficiency and decision-making~\cite{10355694,10092938}. Another significant challenge is the seamless integration of geospatial, geological, and environmental data to construct an accurate and dynamic digital representation of mining operations \cite{zhang2022parallel}. Robotics further contributes to this transformation by facilitating autonomous exploration, real-time monitoring, and hazardous environment assessment, reducing the need for human intervention in high-risk areas ~\cite{bansal2025exploring}. Additionally, safety and risk management are critical considerations, as AI-powered predictive analytics can proactively identify hazardous zones and optimize site operations. However, industry adoption and scalability remain significant hurdles, as mining companies often resist new technologies due to high costs and implementation complexity. Addressing these challenges requires an interdisciplinary approach that leverages AI-driven analytics, robotics, remote sensing, and interactive visualization to drive efficiency and innovation in mining operations.

Generally, the MASTERMINE project proposes a multilayered system architecture that integrates UAVs, geo-referencing technologies, process automation, AI-driven real-time analytics, and immersive virtual environments to revolutionize mining operations. Our approach includes components of the MASTERMINE system, which refer to automated UAV-based data collection, where UAVs equipped with LiDAR and RGB imaging sensors gather detailed geospatial and geological data. Geo-referenced data augmentation is applied through advanced algorithms that enhance data accuracy by mapping UAV-collected information onto existing geospatial frameworks. A continuously updated digital twin of the mining environment enables simulation and predictive analytics while incorporating a Metaverse-based interactive environment that allows stakeholders to analyze, optimize, and make informed decisions in a virtualized mining space.
By leveraging digital twin and Metaverse technologies, MASTERMINE contributes to the scientific community by providing an innovative framework for real-time mining analytics. This approach enhances safety, efficiency, and sustainability by enabling remote monitoring, predictive modeling, and interactive simulations. Furthermore, this research advances the field of digital mining by demonstrating how high-fidelity spatial datasets can improve asset management, workflow automation, and long-term planning in the extractive industry.

Remote sensing based on UAVs is pivotal in building a functional and immersive Metaverse for mining applications. The collected data undergoes geo-referencing and augmentation to create high resolution 3D models of mining sites ~\cite{stothard2023application}. This process facilitates risk assessment and hazard detection by identifying unstable terrain and dangerous zones, optimizes resource extraction through real-time insights into geological formations, supports sustainable mining practices by monitoring environmental impact and waste management, and enables remote training and simulation for personnel, reducing operational risks and improving decision-making ~\cite{nalmpant2025framework}. Additionally, the use of AI-enhanced data processing ensures that the mining Metaverse remains a dynamic and evolving platform, adapting to discoveries and operational requirements ~\cite{du2025industrial}.

The integration of augmented UAV data into Metaverse environments transforms mining operations from traditional, reactive models to predictive, data-driven strategies, ultimately leading to safer, more efficient and sustainable practices. This research underscores the growing importance of digital twins and Metaverse technologies as enablers of Industry 4.0 in the mining sector, offering novel pathways for operational optimization and strategic planning ~\cite{ghahramanieisalou2024digital}.

This work introduces a novel, integrated framework that enhances UAV-based geospatial intelligence in mining operations, addressing critical challenges related to localization accuracy, real-time data synchronization, and operational efficiency. The key contributions of our approach are:
\begin{itemize}
\item 3D Modeling of the Mining Site: We developed a high-precision 3D model of the mining area using LiDAR equipped UAVs to scan the terrain and structures in detail.
\item Object Detection and Projection onto a 3D Map: Using the UAV’s onboard camera and deep learning algorithms, we detect critical objects and project them onto the 3D map, enhancing spatial awareness and situational analysis.
\item Visual Navigation with geo-referenced images: We leverage geo-referenced images for UAV navigation, allowing for more accurate positioning and real-time adaptability to the evolving mining environment.
%\item Near Real-Time Data Synchronization: Our system seamlessly integrates LiDAR scans, camera data, and geospatial references, enabling continuous updates to the 3D model and improving real-time decision-making.
\item Integration into the industrial Metaverse: By linking UAV-acquired data with geospatial intelligence, we create a unified framework that enhances digital twin capabilities for mining operations, supporting more informed and data-driven decision-making.
\end{itemize}
Our approach advances AI-enhanced geospatial analytics in mining, providing a more dynamic, accurate, and real-time representation of mining sites, ultimately improving safety, operational efficiency, and resource management.
The rest of this article is organised as follows. Section II presents previous studies. Section III explains the methodology of our study. Section IV shows the results of implementing the proposed system, while Section V includes the discussion. Finally, Section VI concludes this paper.
\section{Related Work}
The integration of UAVs in mining environments has gained significant attention due to their ability to improve operational efficiency, enhance safety, and enable real-time geospatial analysis ~\cite{asadzadeh2022uav}. UAVs are widely employed in mining applications, including autonomous navigation, object detection, hazard identification, and 3D mapping. These technologies allow mining operations to become safer, more precise, and more efficient, reducing the reliance on manual inspections and improving real-time decision-making capabilities. Despite advancements in UAV-based solutions, challenges remain in areas such as real-time data processing, geospatial accuracy, and interoperability with existing mining infrastructure. Moreover, while current research explores UAV applications in isolated aspects such as mapping or object detection, integrating these capabilities into a cohesive, AI-driven digital twin ecosystem remains challenging. The following sections discuss related studies on UAV navigation, object detection, and 3D mapping in mining environments, highlighting their contributions and limitations.

\subsection{UAV Navigation in Mining Environments}
Effective navigation is a fundamental requirement for the successful operation of UAVs, especially in environments with limited visibility and uncertainty. The ability of a UAV to move accurately and safely in such conditions largely depends on the methods it employs for localization and obstacle avoidance ~\cite{10.1007/978-3-030-34995-0_17}. One of the primary challenges of UAV operations in mining environments is navigation in GNSS-denied areas such as underground mines, tunnels, and heavily occluded open-pit mines. Traditional GNSS-based navigation becomes unreliable in such environments due to signal loss, necessitating alternative methods such as Simultaneous Localization and Mapping (SLAM), Inertial Navigation Systems (INS), and LiDAR-based point cloud mapping \cite{said2021application}.

LiDAR and SLAM-based UAV navigation has been widely used to improve localization accuracy. LiDAR-equipped UAVs scan the environment in real time, generating a point cloud that is used for localization and obstacle detection ~\cite{sarantinoudis2024applications}. However, challenges remain in processing large point cloud datasets efficiently, as well as ensuring accuracy in dynamic mining environments, where terrain changes frequently due to undergoing excavations. AI-driven path planning and obstacle avoidance have also been explored as alternatives. Machine learning algorithms enhance UAV navigation by predicting optimal flight paths and dynamically adjusting navigation parameters based on real-time sensor data ~\cite{elbazi2023digital}. Even though promising, AI-driven navigation systems often require high computational resources, limiting their real-time applicability in complex sites. Some approaches use computer vision techniques to aid UAV navigation, relying on stereo cameras and optical flow algorithms to estimate movement and detect obstacles ~\cite{uddin2023landing,papapetros2022visual}. However, dust, low lighting, and occlusions in underground mines often reduce the effectiveness of purely vision-based navigation.

Unlike previous research that focuses on isolated navigation solutions, our approach integrates AI-driven visual positioning technique, LiDAR mapping, and deep learning-based object detection, enabling UAVs to adapt dynamically to evolving mining environments while maintaining high localization accuracy.

\subsection{Object Detection and Hazard Identification in Mining}
Object detection is a critical component of UAV-based mining operations, enabling the identification of structural instabilities, loose rocks, underground voids, and worker movements. Most existing research leverages computer vision and deep learning models for near real-time object detection. Such approaches, particularly Convolutional Neural Networks (CNNs) and YOLO-based (You Only Look Once) object detection models, have been widely applied to UAV imagery in identifying hazardous zones and structural anomalies ~\cite{cranford2023conceptual,zhang2024rotator}. However, these methods often require large training datasets and struggle with variability in mining conditions, such as changing lighting, dust interference, and occlusions. Some studies have leveraged multispectral and hyperspectral imaging to detect variations in rock composition, mineral deposits, and potential hazard zones. These advanced imaging techniques allow for detailed material classification, which can help identify weak rock structures before they collapse \cite{elbazi2023digitalb}. However, high data processing requirements and the need for specialized calibration present challenges for real-time deployment.

Additionally, IoT-based ground sensors are increasingly being used alongside UAVs to provide comprehensive hazard assessments. These sensors measure seismic activity, temperature changes, and gas emissions, complementing UAV-based visual inspections ~\cite{singh2022applications}. Despite their potential, integrating these diverse datasets into a real-time, interactive hazard monitoring system remains a challenge.
Recent advancements have explored innovative approaches to enhance UAV-based mining operations. A light-weight approach for safe landing in populated areas has been proposed, which optimizes UAV navigation and emergency landing decisions ~\cite{mitroudas2024light,10355707}. This method improves UAV operational safety by leveraging real-time environmental data to identify suitable landing zones, reducing risks in densely populated or hazardous areas. Furthermore, enhancing satellite semantic maps with ground-level imagery has been proven advantageous for spatial awareness for UAV-based object detection applications  ~\cite{balaska2021enhancing}. By integrating satellite data with UAV-captured images, researchers can enhance the resolution and accuracy of detected hazards, providing a more comprehensive mapping solution for mining operations ~\cite{nguyen2022application}.

The combination of UAV-based imagery, ground sensors, and advanced mapping techniques is crucial for developing robust hazard detection systems. Recent advances in computer vision introduce novel architectures such as Vision Transformers (ViTs) and diffusion models, which show strong performance in object detection and segmentation tasks under challenging conditions ~\cite{li2024transformer, he2025diffusion}. These methods can potentially address some of the limitations conventional CNN-based models have in challenging environments, particularly under variable lighting, occlusions, or class imbalance. 

\subsection{3D Mapping in Mining}
The mining industry heavily relies on precise spatial data to optimize operations, assess environmental impact, and improve safety. UAV-based 3D mapping has emerged as a powerful tool in this domain, providing high-resolution and cost-effectiveness, as well as efficient solutions for terrain analysis, volumetric assessments, and mine planning. These technologies enable mining missions to monitor excavation progress, detect structural instabilities, and improve decision-making through accurate geospatial information ~\cite{doi:10.1080/25726668.2020.1786298}.
Among the most widely adopted techniques for 3D mapping in mining are LiDAR and photogrammetry. LiDAR technology utilizes laser pulses to measure distances with high precision, generating dense point clouds that accurately represent the topography of open-pit and underground mining environments. This method is particularly useful for detecting fine geological features, assessing surface deformations, and creating Digital Elevation Models (DEMs). However, LiDAR-based mapping requires expensive hardware and complex data processing workflows, making it less accessible for smaller mining operations ~\cite{Azhari2017A}.

On the other hand, photogrammetry-based mapping relies on high-resolution images captured by UAVs, which are then processed using advanced computer vision algorithms to reconstruct 3D models. This technique is more cost-effective than LiDAR and can cover large areas quickly. Despite its advantages, photogrammetry often suffers from a reduced precision in featureless or low-texture environments, such as underground mines, where lighting conditions and occlusions can affect the quality of the data ~\cite{jozkow2021monitoring}.
While both LiDAR and photogrammetry provide highly detailed representations of mining sites, they typically require extensive post-processing and lack real-time adaptability to site changes. The integration of real-time data collection and processing is a growing research area, aiming to enhance the responsiveness of 3D mapping technologies in mining operations. Advances in edge computing and AI-driven data fusion techniques are expected to reduce processing time and enable near-instantaneous analysis of spatial data, offering new possibilities for automation ~\cite{elbazi2023digital}. 

\subsection{Integration of 3D Mapping into Digital Twins}
Digital twins take 3D mapping a step further by providing near real-time, interactive virtual environments replicating mining operations ~\cite{to2021drone}. Some studies propose integrating UAV data into near real-time digital twin frameworks, allowing continuous updates of mining models and providing enhanced simulation capabilities for mine planning and predictive maintenance ~\cite{sarantinoudis2024applications}. However, many digital twin implementations lack interoperability with existing mining software and fail to integrate multi-source data streams effectively. Emerging research explores using Metaverse technologies to create immersive, interactive digital replicas of mines for stakeholder engagement and training ~\cite{akbulut2025review}. Despite their potential, current solutions often lack real-time data synchronization and struggle with scalability in large mining operations.
In addition to academic initiatives, several large-scale industrial deployments highlight the global interest in digital transformation of mining. For instance, the Rio Tinto's ``Mine of the Future'' program integrates autonomous UAVs and AI-based analytics for real-time monitoring and decision support across its operations in Australia\footnote{https://www.riotinto.com}. Similarly, Vale has piloted digital twin platforms in Brazilian mines for predictive maintenance and hazard identification using UAV-collected data\footnote{https://www.worldsensing.com/knowledge-center/digital-transformation-mines-insights-from-vale/}. These initiatives underscore the growing industrial momentum towards integrated digital ecosystems in mining, aligning with the objectives of our proposed framework.
\subsection{Motivating the Proposed Approach}
The proposed system differs from prior work ~\cite{10017382} that addresses UAV-based mapping or object detection in isolation by providing an integrated framework that combines geo-referenced vision-based localization, LiDAR terrain modeling, and deep learning-based object recognition. This architecture enables accurate spatial mapping of mining environments and supports structured data export for downstream integration into digital twin platforms. Rather than focusing on real-time autonomous navigation, our approach emphasizes modularity, data fidelity, and compatibility with industrial geospatial systems. UAV-acquired data is processed post-flight and aligned with site-specific geospatial references, facilitating spatial situational awareness and infrastructure monitoring. By linking AI-driven vision processing with precise geo-localization techniques, the system contributes to safer, more efficient, and data-informed mining workflows.
\begin{figure}[htbp]
\centerline{ \includegraphics[width=0.5\textwidth]{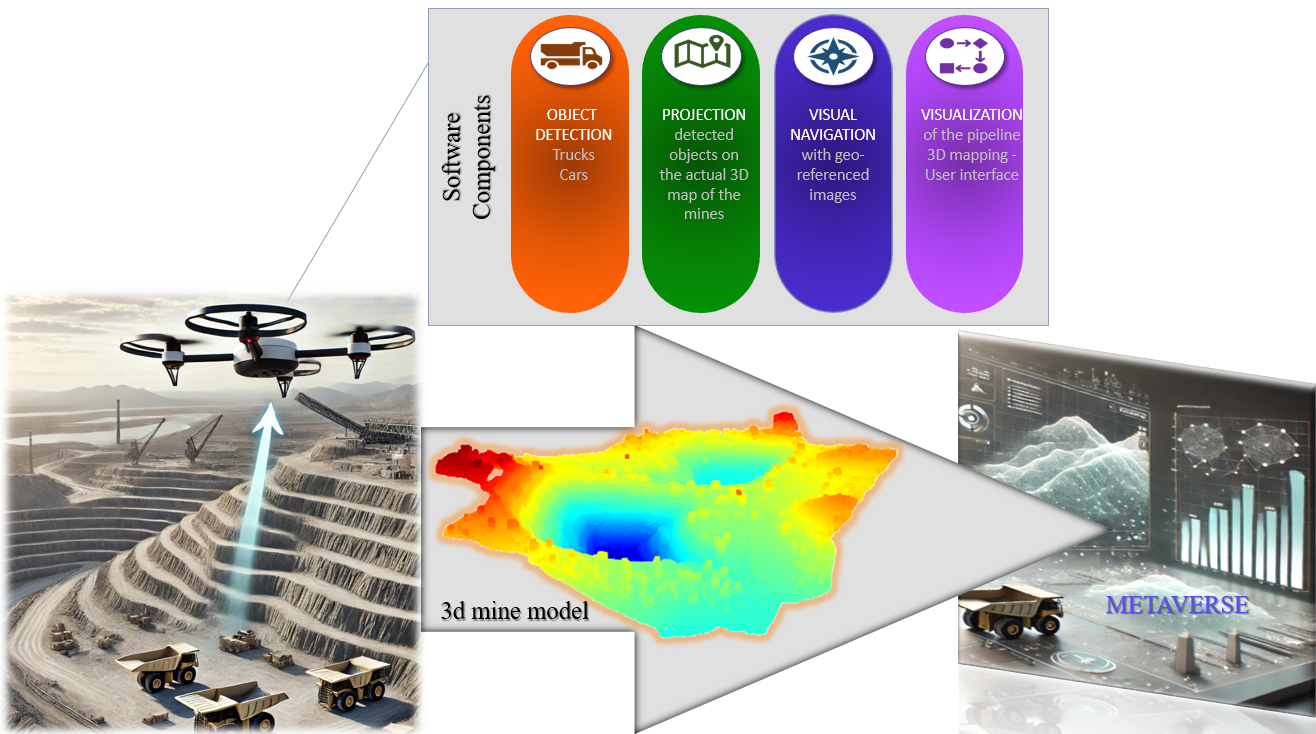}}
\caption{Data pipeline from UAV-based mine scanning to Metaverse visualization. A DJI Matrice 350 RTK collects topographic data of an open-pit mine (left), which is processed into a 3D terrain model (center). Software components support object detection, georeferenced projection, and user interaction. The resulting digital twin is visualized in a virtual workspace (right), enabling remote analysis and decision support.}
\label{fig:flowPAPER}
\end{figure}

\begin{figure}[htbp]
\centerline{ \includegraphics[width=0.5\textwidth]{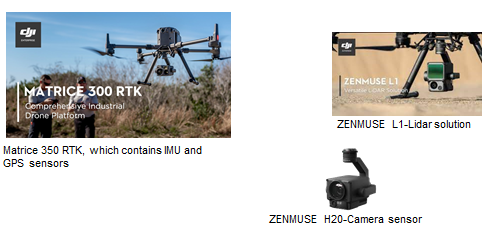}}
\caption{The equipment used to develop the proposed system.}
\label{fig:equip}
\end{figure}

\begin{figure}[htbp]
\centerline{ \includegraphics[width=0.5\textwidth]{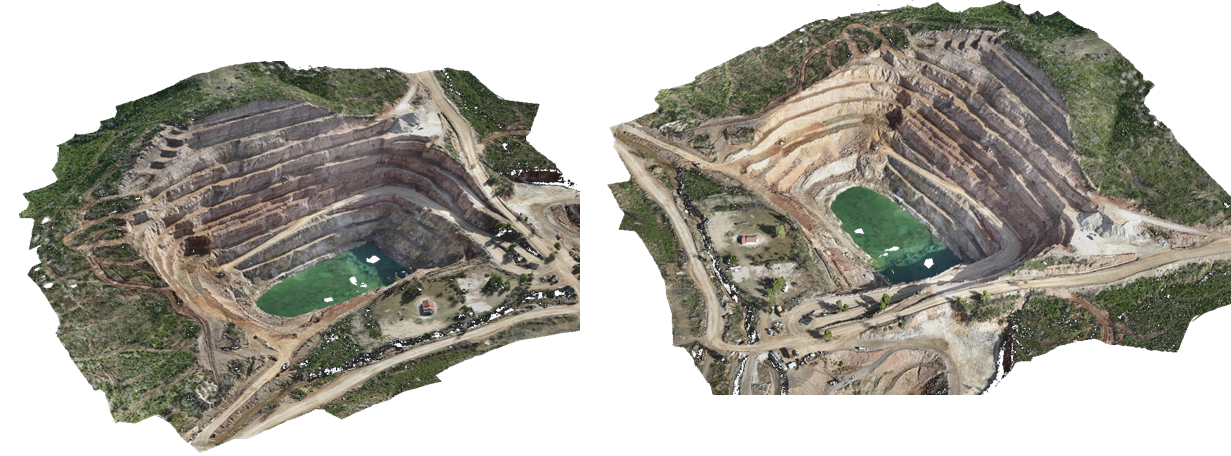}}
\caption{TERNA MAG test site: A 3D representation of Bompaka's open pit.}
\label{fig:terna}
\end{figure}

\section{Proposed Methodology}
\subsection{System Overview}
The proposed system contributes to a Metaverse-driven mining monitoring framework by enabling UAV-based geo-referenced data acquisition for object detection and 3D positioning in industrial environments (Fig. \ref{fig:flowPAPER}). Through the integration of UAV-collected sensor data with digital terrain models, the system supports accurate spatial mapping and enhanced situational awareness in open-pit mining contexts.

Rather than implementing real-time communication with the Metaverse platform, the presented system delivers processed outputs, such as geolocated object annotations and 3D reconstructions, via standardized formats to external actors responsible for virtual environment integration. This forms the basis of an augmented digital twin, where discrete mission outputs are synchronized with a virtual representation of the mining site to improve operational monitoring and safety analysis.

The architecture comprises three main components: the UAV platform, the on-ground computational infrastructure, and the software pipeline that transforms raw sensor inputs into geo-referenced spatial products. In our deployment, a DJI Matrice 350 RTK UAV equipped with a Zenmuse L1 LiDAR and a Zenmuse H20 RGB camera was utilized (Fig.~\ref{fig:equip}), with GNSS and IMU aiding precise geo-positioning and motion tracking.

The computational infrastructure combines edge computing and cloud computing. On-board processing optimizes bandwidth and reduces latency, while cloud-based AI handles object detection and 3D reconstruction. The processed data are integrated into GIS-based 3D models, improving spatial awareness in the mining metaverse. At the software level, a YOLOv8-based deep learning model detects and classifies mining objects, while 3D point cloud augmentation fuses UAV imagery with LiDAR data for precise localization. A vision-based positioning module ensures spatial consistency and an interactive 3D visualization interface provides near real-time geo-referenced object mapping for better decision-making. This system improves situational intelligence in mining, optimizing workflows, efficiency, and safety in industrial operations.

\subsection{Data Collection}
The data collection process is a critical step in the development of a robust geo-referenced object detection and positioning system within the mining Metaverse framework. 
The proposed methodology involves the deployment of a Matrice 350 RTK UAV, equipped with state-of-the-art sensors to capture high-precision environmental data from open-pit mining sites. 
The TERNA MAG Open Pit Mine serves as the primary test site for data acquisition (Fig.~\ref{fig:terna}).

The Zenmuse L1 LiDAR sensor is utilized to generate a detailed 3D point cloud of the mining environment, capturing terrain morphology with high accuracy.
Additionally, the Zenmuse H20 RGB camera provides high-resolution imagery, facilitating both object detection and 3D spatial mapping. Finally, measurements from onboard IMU and GNSS RTK receiver are combined with the above collected data to achieve high-precision geo-referencing.

Three primary data types are collected:
\begin{itemize}
\item RGB images containing targeted object classes, as provided by TERNA.
\item LiDAR point cloud data of the open-pit mine, used for terrain reconstruction and geospatial modeling.
\item Geo-referenced RGB images, enabling accurate localization and integration within the 3D mining Metaverse.
\end {itemize}
The data collection process follows a structured methodology. 
UAV flights are carefully planned to ensure comprehensive coverage of the mining site. 
The collected RGB images are annotated to train object detection models, while the LiDAR point cloud is processed to create a high-fidelity 3D reconstruction of the terrain. 
In addition, geo-referenced images are generated to support our vision-based positioning approach, where unregistered UAV images can be matched to the existing dataset for accurate localization.
The acquired dataset serves as the foundation for object detection, 3D mapping, and vision-based UAV positioning, which are integral components of the MASTERMINE Metaverse system. 
The seamless integration of these data sources enhances situational awareness, operational efficiency, and near real-time decision-making in mining environments.

The UAV platform used in this study was the DJI Matrice 350 RTK, selected for its robust flight stability, long endurance (up to 55 minutes), and compatibility with multiple payloads. For sensor integration, two modules were employed: the Zenmuse H20, a multi-sensor RGB camera with zoom and laser rangefinder capabilities, and the Zenmuse L1, a LiDAR module featuring an integrated IMU and a visible-spectrum camera for 3D point cloud generation.

This sensor configuration enabled high-resolution image acquisition and accurate 3D spatial mapping under diverse environmental conditions. Although the hardware setup was determined by project availability and logistical constraints, it was well-suited for the tasks of object detection, geo-referenced localization, and digital terrain reconstruction. The optical and geometric characteristics of the H20 and L1 were taken into account during flight planning to ensure optimal image overlap, LiDAR coverage, and precise alignment with spatial reference models.

\subsection{Object Detection and Projection}\label{sec:objdet}
In this study, object detection was performed utilizing the YOLOv8  architecture~\cite{redmon2016you}, specifically the YOLOv8l (large) variant provided by the Ultralytics framework. 
The large model was selected due to its superior detection accuracy and enhanced feature extraction capacity, which are critical for mining site monitoring tasks, especially in complex aerial imagery. 
This choice ensured reliable detection performance, particularly for small or visually similar objects, where lightweight models may underperform.
The initial training phase employed a proprietary dataset sourced from TERNA’s MAG mining operations, comprising aerial imagery collected via custom UAVs.
To validate the generalizability of the detection model, a secondary dataset was curated using the Matrice 350 RTK UAV.
This dataset contains 804 images under diverse operational scenarios. 
Annotation was performed through the Roboflow platform\footnote{https://roboflow.com}, ensuring standardized labelling across all classes.
The dataset consists of eight object classes critical to mining site monitoring, i.e ~\textit{ bulldozer} (229 samples), ~\textit{car} (327 samples), ~\textit{driller} (254 samples), ~\textit{dump truck} (675 samples), ~\textit{excavator} (384 samples), ~\textit{grader} (63 samples), ~\textit{human} (245 samples), and ~\textit{truck} (214 samples).
Additionally, the preprocessing steps included automated image orientation correction to normalize input alignment, tiling of each image into a 2×2 grid (four tiles per image) to enhance object detection granularity.
To enhance model robustness against illumination variability, grayscale transformation was applied to $15\%$ of the dataset images.
Furthermore, data augmentation was employed to mitigate overfitting and improve the model’s generalization capability. 
Lastly, a random rotation of $90$ degrees was applied during the preprocessing phase.
This augmentation strategy aimed to simulate various real-world conditions encountered in aerial surveillance, such as differing altitudes, angles, and lighting conditions.
The resulting output dataset has a size of 5592 images, and it is partitioned into training ($85\%$), validation ($9\%$), and test ($6\%$) sets, corresponding to 4752, 516, and 324 images, respectively.
The YOLOv8l model was initialized with pretrained weights to leverage knowledge transfer from large-scale generic datasets. 
To this end, training was conducted using the Ultralytics framework following the hyperparameter configuration as presented in Table~\ref{tab:params}, , on both the TERNA MAG and the custom Matrice 350 RTK datasets.
\begin{table}
    \centering
    \caption{Training parameters}
	\setlength{\tabcolsep}{4pt}
    \begin{tabular}{c c}
    \hline
       Parameters & Values \\
       \hline
        Epochs & 200\\
        Batch size & 64\\
        Input image resolution & 640 × 640 pixels \\
        Augmentation scaling factor &  0.4\\
        Rotation range & 360 degrees \\     
    \end{tabular}
    \label{tab:params}
\end{table}
\begin{figure*}[!t]
    \centering
    \subfloat{%
        \includegraphics[width=0.36\textwidth]{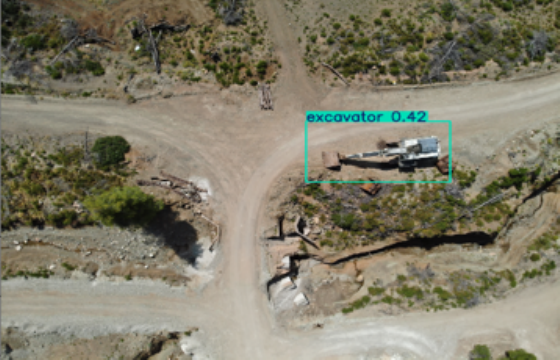}%
        \label{subfig:detobj}}
    \hspace{0.05\textwidth}
    \subfloat{%
        \includegraphics[width=0.45\textwidth]{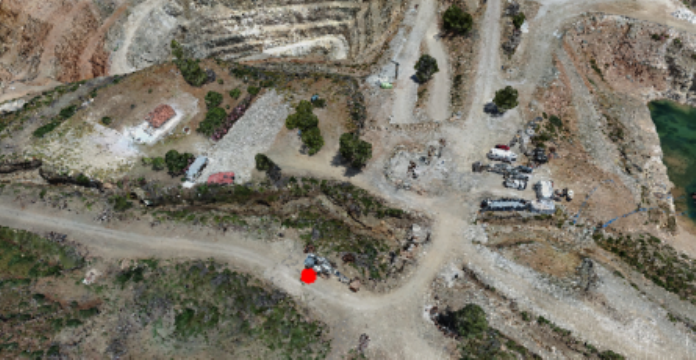}%
        \label{subfig:3dobj}}
    \caption{(left image) The corresponding detected objects are visualised in the 2D image captured by the UAV.(right image) Projection of detected objects onto the 3D mine environment, highlighting their spatial positioning within the point cloud. }
    \label{fig:proj}
\end{figure*}
Following object detection, the spatial positioning of the detected objects within the 3D map of the mining environment was achieved by geometric projection techniques. 
The objective was to accurately localize detected objects within a unified geospatial reference frame by projecting $2D$ image detections onto a registered 3D point cloud. The internal camera parameters, focal length $(f_x, f_y)$, and principal point $(c_x, c_y)$ were extracted from the image metadata and further refined through camera calibration, forming the intrinsic matrix $K$ used in the pinhole camera model.

The acquired image metadata also provided the camera’s geolocation (latitude, longitude, altitude) and its orientation in the North-East-Down (NED) reference frame, expressed as roll, pitch, and yaw angles. Since our global reference system was UTM, which follows an East-North-Up (ENU) convention, we converted the camera orientation from NED to ENU by applying a fixed coordinate transformation.

To reconcile the orientation of the image plane with the global frame, we also had to account for differences between the standard camera coordinate system (X: right, Y: down, Z: forward) and the NED convention. This was achieved by applying a fixed rotation, aligning the internal camera axes to match the NED orientation given in the metadata. The full rotation matrix defining the transformation from the camera to the global ENU frame was constructed by chaining: (i) the rotation from NED Euler angles, (ii) the NED-to-ENU frame conversion, and (iii) a camera-to-NED alignment matrix that remaps the physical camera axes into the NED orientation space.

To visualize the spatial correspondence between image detections and the 3D scene, the point cloud was projected into the image frame using the camera’s extrinsic and intrinsic parameters. Each 3D point was first transformed into the camera frame using the full pose matrix, then projected into 2D pixel coordinates via the intrinsic matrix $K$, following the standard pinhole model. Points falling outside the image bounds or behind the camera were discarded. The remaining valid projections revealed which parts of the 3D scene were visible from the current view and enabled spatial association with detected objects by checking proximity to the projected bounding boxes.
Figure~\ref{fig:proj} presents an example of the projection process.

\subsection{Robust Vision-Based UAV Localization in GNSS-Denied Mining Environments}
\begin{figure}[htbp]
\centerline{ \includegraphics[width=0.5\textwidth]{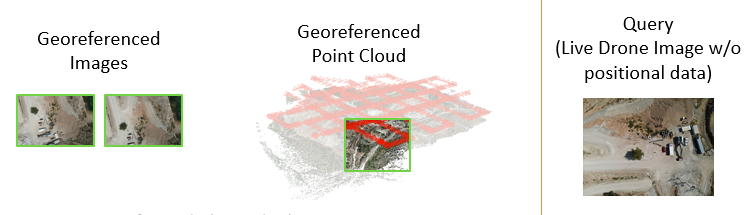}}
\caption{A live query image without GNSS data is associated with a specific region of the geo-referenced point cloud based on visual similarity with reference images. }
\label{fig:visual_matching}
\end{figure}

\begin{figure}[htbp]
\centerline{ \includegraphics[width=0.5\textwidth]{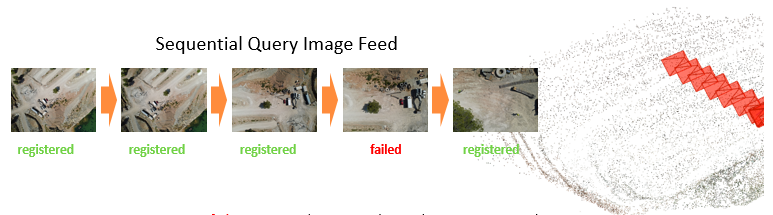}}
\caption{Sequential query images are processed for position estimation. While some frames may fail due to unmapped regions or noise, successfully registered images enable trajectory alignment and improved localisation through relative positioning.}
\label{fig:GEO3}
\end{figure}

The UAV vision positioning framework combined visual place recognition, geometric pose estimation, and Structure-from-Motion (SfM) to enable GNSS-free localization. %~\cite{TSAMPIKOS}. 
A geo-referenced 3D model was first constructed using COLMAP~\cite{schoenberger2016sfm} from a set of database images with embedded GNSS metadata, captured over the target area. SIFT~\cite{lowe2004distinctive} features were extracted and matched using spatial priors based on GNSS data, producing a globally aligned sparse reconstruction that served as the localization reference.
\begin{figure}[htbp]
\centerline{ \includegraphics[width=0.5\textwidth]{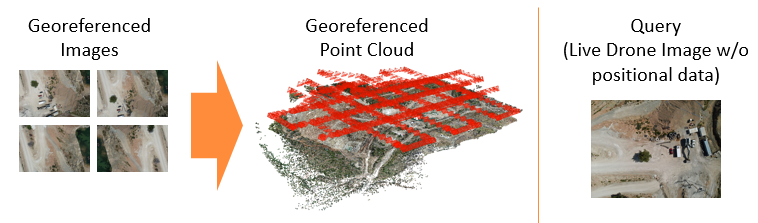}}
\caption{Geo-referenced reconstruction and visual localisation pipeline: Geo-referenced images are used to build a 3D point cloud, which serves as a reference for localising query images captured without positional data.}
\label{fig:GEO}
\end{figure}
Query images, captured without GNSS, were initially matched to the geo-referenced model using a global descriptor-based visual place recognition method as shown in Fig.~\ref{fig:GEO}. 
As illustrated in Fig.~\ref{fig:visual_matching}, the query image is visually matched to a subset of geo-referenced images and their corresponding region within the 3D model. 
For each retrieved database image, 2D–3D correspondences were obtained using COLMAP’s SIFT features, and a Perspective-n-Point (PnP) solver was used to recover the full 6-DoF pose of the query image.
However, when this registration failed, typically due to insufficient feature matches or occlusions, a secondary SfM model was constructed from the query sequence itself as presented in Fig.~\ref{fig:GEO3}.
This non-geo-referenced reconstruction was built using COLMAP with SIFT feature extraction capped at 10,000 features per image. Exhaustive matching and incremental mapping produced a locally consistent but unscaled and unreferenced trajectory of the query sequence. To anchor this local trajectory to the global frame, a similarity transform was estimated using the subset of query images that had been successfully registered via visual place recognition and PnP. This final alignment step allowed the full query sequence, including previously failed images, to be expressed within the global UTM coordinate frame.

\subsection{Holistic Visualization}
\begin{figure*}[htbp]
\centerline{ \includegraphics[width=0.9\textwidth]{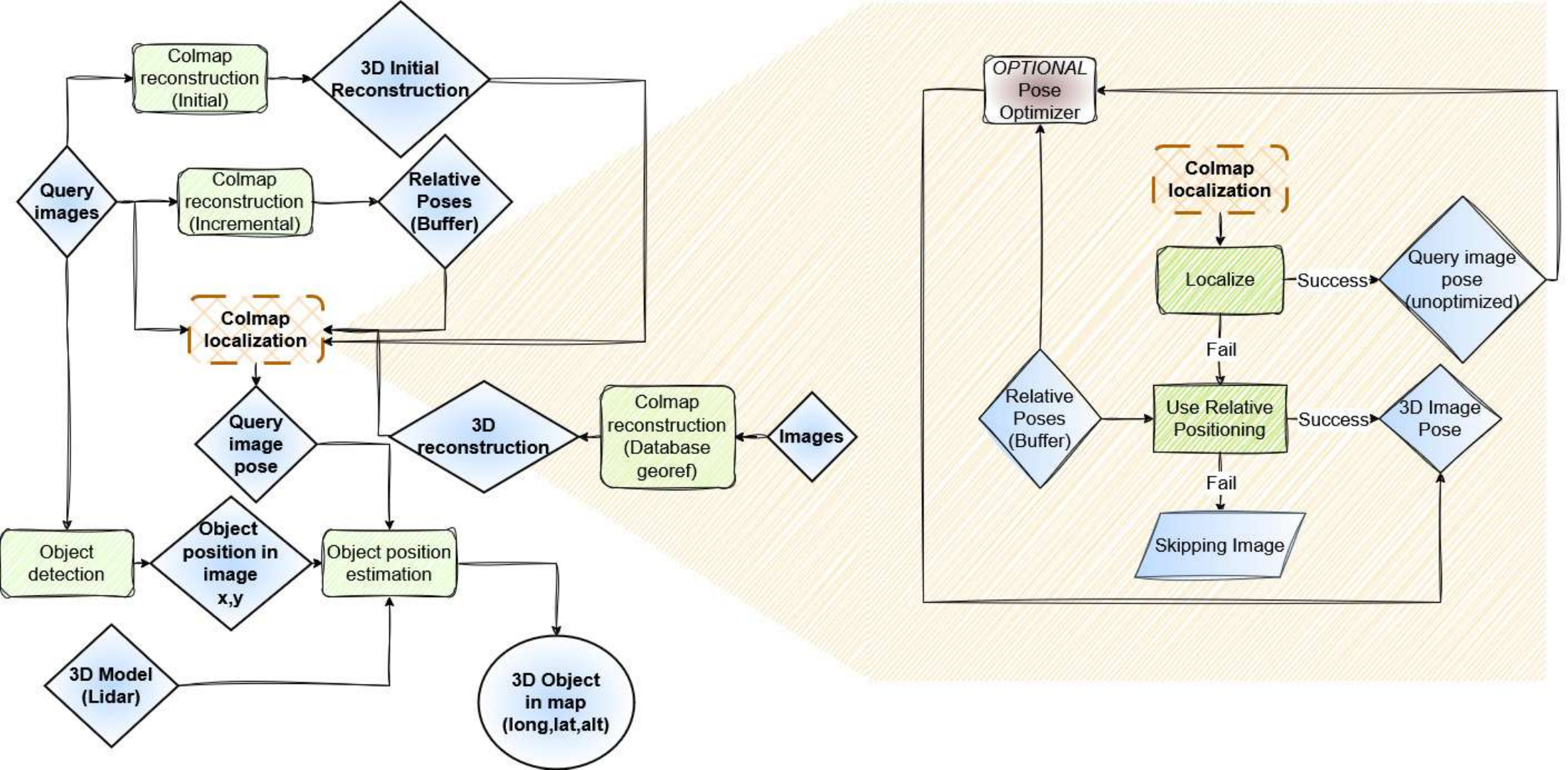}}
\caption{Overview of the proposed UAV vision positioning and object detection framework. }
\label{fig:blockdiag}
\end{figure*}

Our complete implementation is presented in Fig.~\ref{fig:blockdiag}.
The framework integrates 3D reconstruction, visual localization, and object detection and projection into a cohesive pipeline, enabling accurate geospatial positioning and mapping of detected objects within mining environments. As presented in Fig.~\ref{fig:blockdiag}, the overall workflow of the proposed system is delineated through a modular block diagram. 
In this representation, the blue blocks denote data structures or intermediate outputs, while the green blocks correspond to implemented processes or functional modules.
The COLMAP localization component, which plays a pivotal role in the UAV's vision-based positioning, is further elaborated in the detailed subdiagram on the right side of the Fig.~\ref{fig:blockdiag}, outlining the internal logic and decision-making process of the localization pipeline.
The system comprises of three basic stages: i) the 3D reconstruction processes (both non-geo-referenced and geo-referenced) that provide the spatial frameworks, ii) the visual localization via COLMAP, with conditional fallback to relative positioning if direct localisation fails, and iii) the object detection and subsequent 3D object positioning, which projects detections from 2D image coordinates onto the global 3D map. The complete implementation was evaluated in simulation for a navigation trajectory, with our custom near real-time data. 
Lastly, the complete implementation of our work is available on our repository\footnote{https://github.com/duth-lra/mastermine}.
\subsection{Integration into the Industrial Metaverse}

The proposed system was designed to generate geo-referenced object detection outputs and 3D localization data suitable for integration into an industrial Metaverse platform. In the context of the TERNA MAG deployment, the system produced orthorectified imagery and annotated 3D spatial data, which were exported via standardized formats (e.g., GeoJSON, PLY, and CSV) and delivered to the platform operator.

The Metaverse platform itself was developed and maintained by a separate project partner, and our role focused on providing structured geospatial inputs from aerial sensing operations. Although the overall architecture supports near real-time workflows, the current implementation operates in a post-flight, batch processing mode. Object detection and 3D reconstruction were conducted offline, and data integration into the Metaverse environment was performed asynchronously through ingestion interfaces provided by the platform developers. This approach enabled accurate spatial visualization and object presence tracking within the digital twin of the mining site, supporting enhanced situational awareness without requiring continuous connectivity or onboard computation.

\section{Experimental Results}
\subsection{Object Detection and Positioning in Real Environments}
The performance of the object detection and positioning system was evaluated using a dataset of 5592 annotated UAV-captured images from mining environments, as presented in Section~\ref{sec:objdet}. The YOLOv8l model, chosen for its superior accuracy and capacity, was trained for 200 epochs with a combination of data augmentation techniques (viz., grayscale transformation, random rotation, and image tiling), ensuring model robustness against real-world aerial imaging conditions.
\begin{figure*}[!ht]
    \centering
    \includegraphics[width=\textwidth]{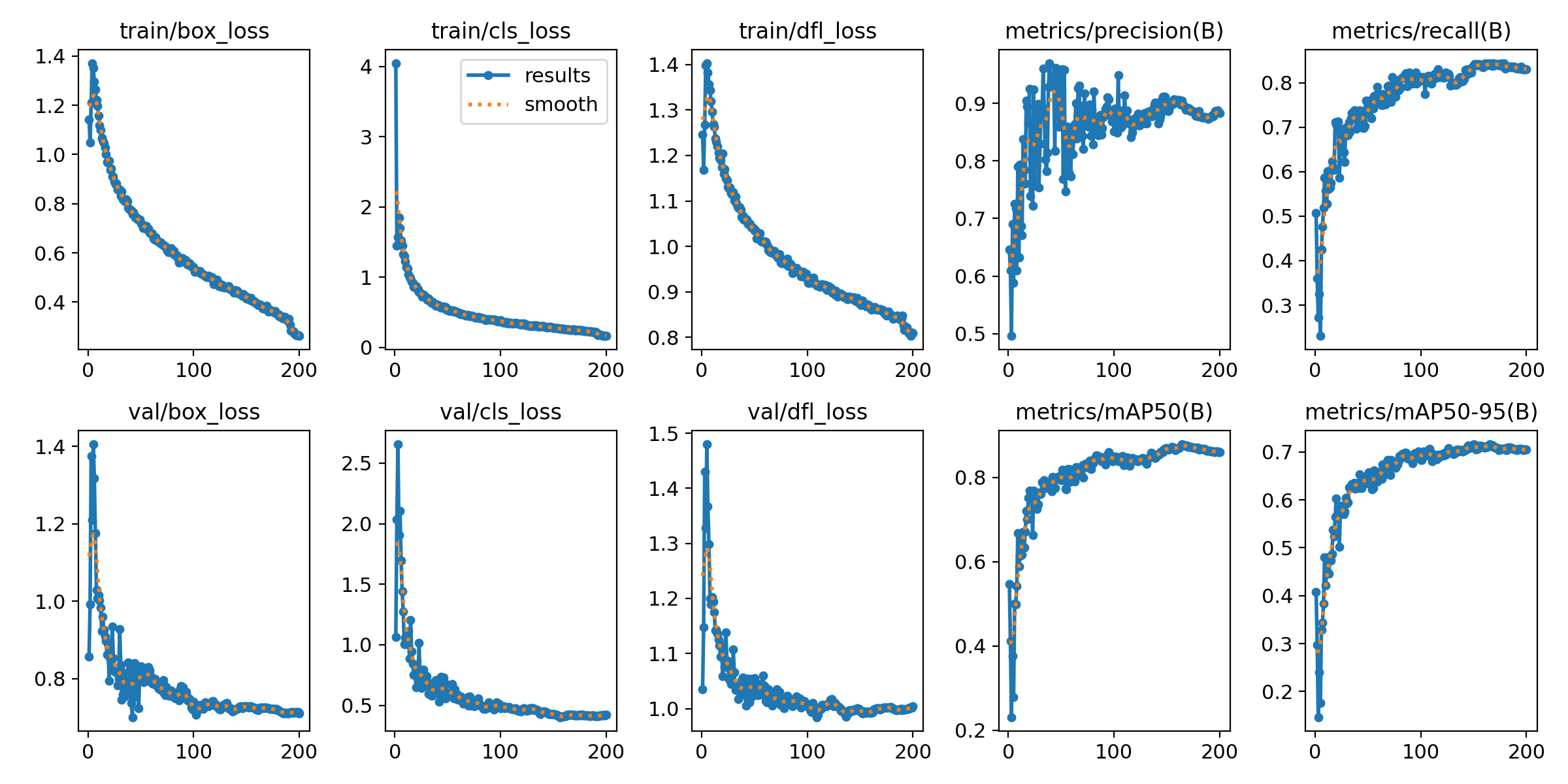}
    \caption{Training and validation curves for the YOLOv8l model. Top row: training loss components. Bottom row: validation metrics including precision, recall, mAP@0.5 and mAP@0.5:0.95.}
    \label{fig:training_curves}
\end{figure*}

Quantitative results indicate a mean Average Precision (mAP) of \textbf{81.4\%} at an IoU threshold of 0.5, and \textbf{63.2\%} at IoU=0.5:0.95. Final evaluation also yielded \textbf{92\% precision} and \textbf{78.4\% recall}, demonstrating the model’s reliability in identifying key operational and safety-related objects such as \textit{trucks}, \textit{excavators}, and \textit{human} workers. These performance metrics are visualized in Fig.~\ref{fig:training_curves}, which also shows convergence in training and validation loss components across epochs.

\begin{figure}[htbp]
    \centering
    \includegraphics[width=0.45\textwidth]{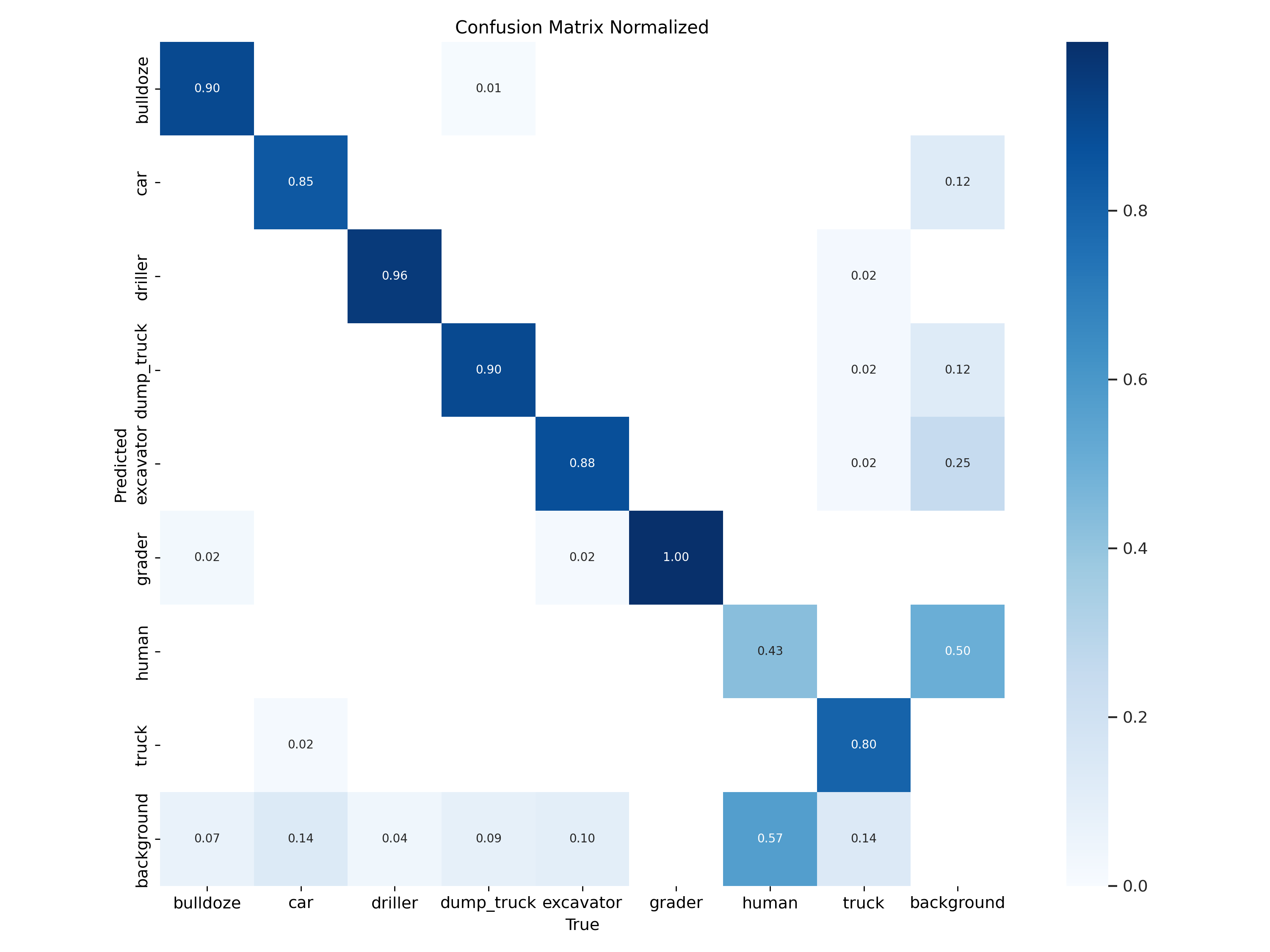}
    \caption{Normalized confusion matrix of the YOLOv8l model across eight object classes. Highest confusion observed amaong \textit{human}, \textit{truck}, and \textit{background} labels.}
    \label{fig:confusion_matrix1}
\end{figure}

A normalized confusion matrix (Fig.~\ref{fig:confusion_matrix1}) further highlights class-wise performance. Most industrial machinery classes, including \textit{driller}, \textit{bulldozer}, and \textit{dump truck}, exceed 85\% classification accuracy. However, the \textit{human} class displayed considerable confusion with the \textit{truck} and \textit{background} labels, with 50\% and 43\% misclassification rates respectively, reflecting visual ambiguity in aerial views.

\begin{figure}[htbp]
    \centering
    \includegraphics[width=0.45\textwidth]{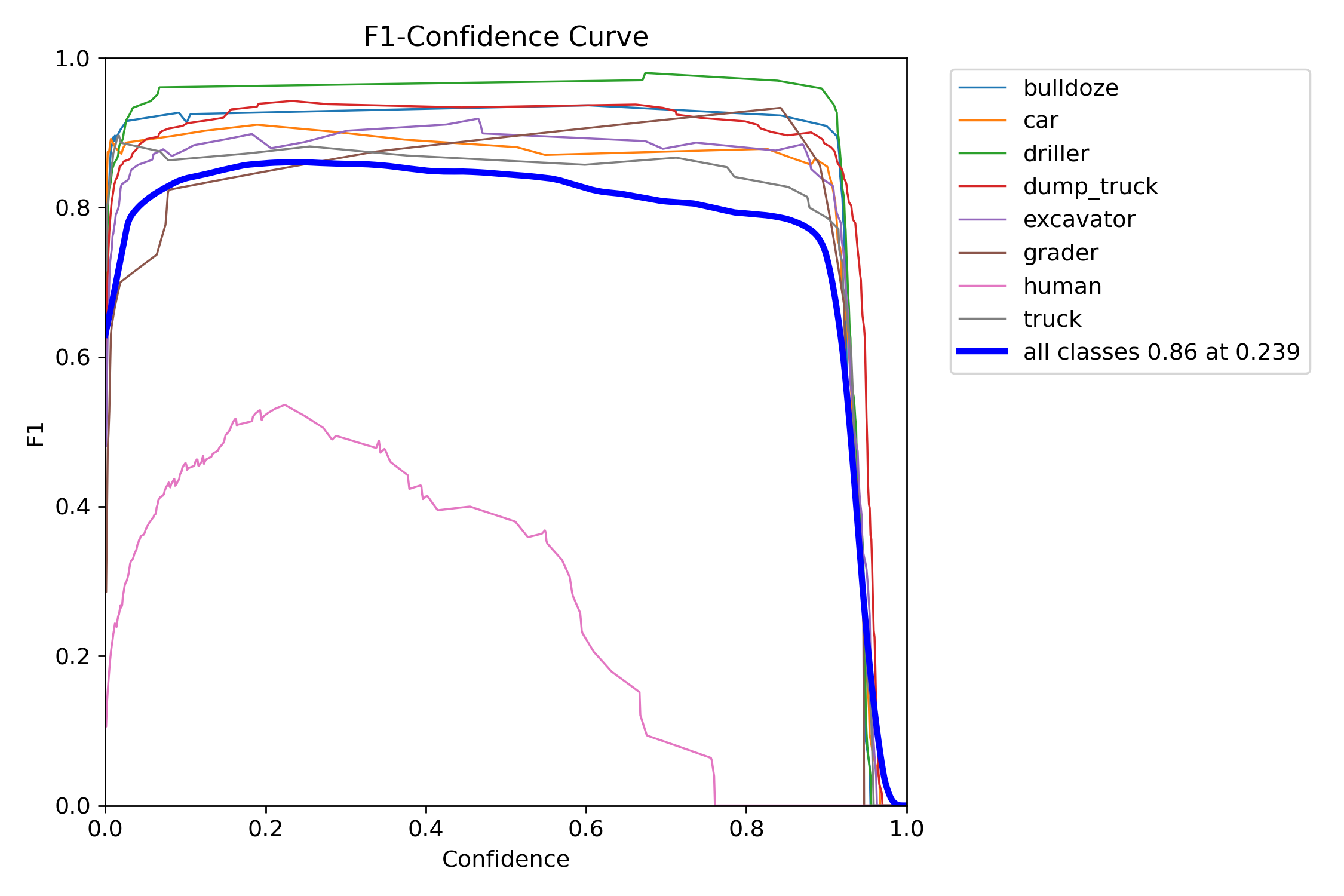}
    \caption{F1-confidence curves for each class. The global optimum (bold line) reaches an F1 score of 0.86 at confidence 0.239.}
    \label{fig:f1_curve}
\end{figure}

To better understand model behavior across different confidence levels, F1-confidence curves were analyzed for each class (Fig.~\ref{fig:f1_curve}). These show that while most object classes achieve F1 scores above 0.9 over a wide confidence range, the \textit{human} class demonstrates significantly lower F1, peaking below 0.55. A global confidence threshold of \textbf{0.239} was identified, yielding a maximum averaged F1 score of \textbf{0.86} across all classes.

In addition to classification, detected objects are spatially positioned in a reconstructed 3D environment through geometric projection. Each detection is accurately mapped to its corresponding location in the point cloud using intrinsic and extrinsic camera parameters. This facilitates spatial tracking and enhances situational awareness within the mining Metaverse. The fusion of RGB imagery with LiDAR point clouds enables semantic-rich digital twins of the environment.

\begin{table}[!ht]
\caption{Summary Statistics for Translation and Orientation Localization Errors}
\label{tab:error_summary}
\centering
\begin{tabular}{|l|c|}
\hline
\textbf{Metric} & \textbf{Value} \\
\hline
Mean Translation Error (m) & 3.69 \\
Standard Deviation Translation Error (m) & 1.79 \\
Mean Orientation Error ($^\circ$) & 1.77 \\
Standard Deviation Orientation Error ($^\circ$) & 0.77 \\
\hline
\end{tabular}
\end{table}

\begin{figure}[htbp]
\centering
\includegraphics[width=0.45\textwidth]{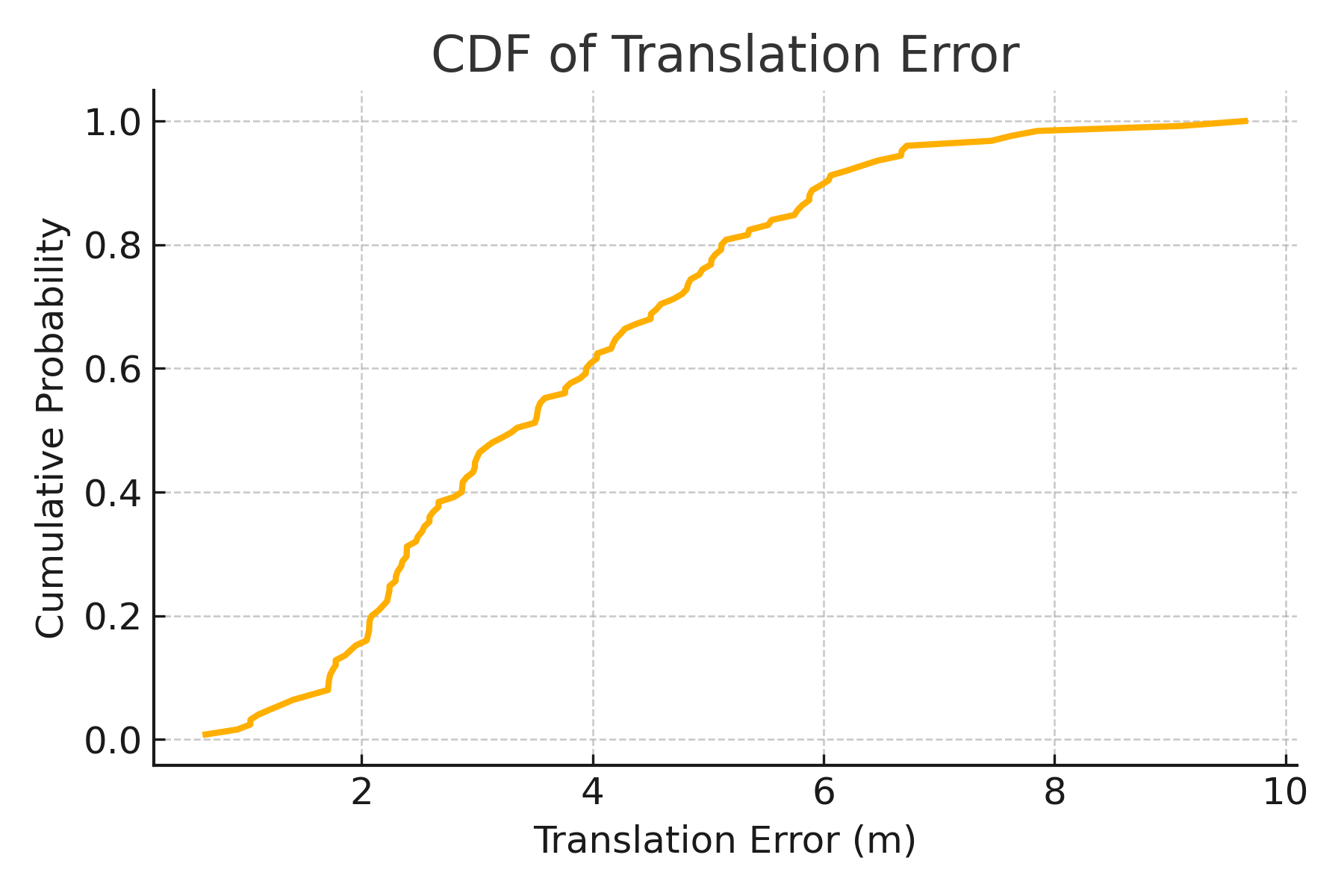}
\caption{Cumulative Distribution Function (CDF) of translation localisation error across all test queries.}
\label{fig:cdf_translation}
\end{figure}

\begin{figure}[htbp]
\centering
\includegraphics[width=0.45\textwidth]{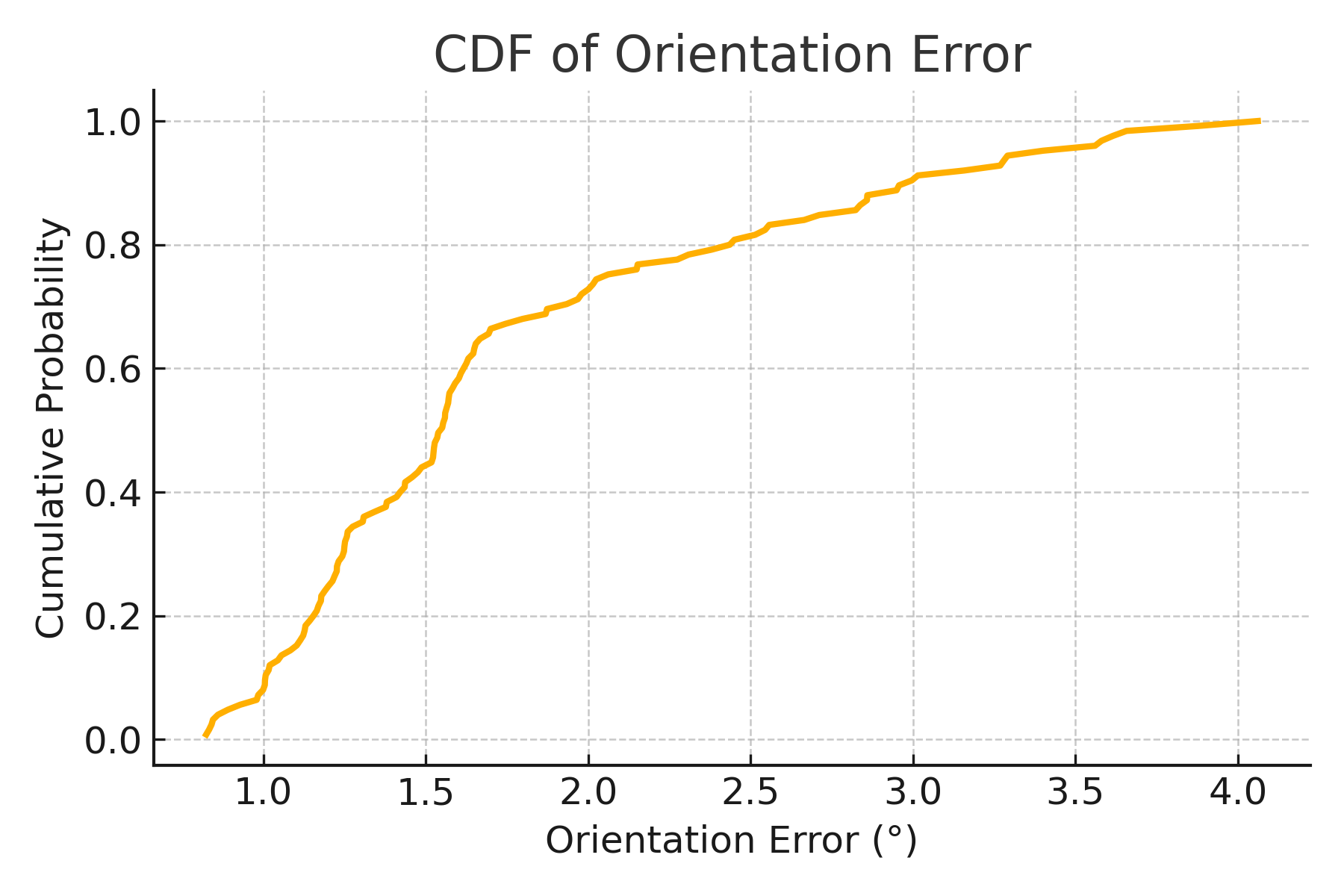}
\caption{Cumulative Distribution Function (CDF) of orientation localisation error across all test queries.}
\label{fig:cdf_orientation}
\end{figure}
\subsection{Vision-Based UAV Positioning Performance}%System's Performance}
% The localization performance of the proposed system was evaluated based on translation and orientation error metrics, following established 6-DoF outdoor localisation benchmarks ~\cite{sattler2018benchmarking}. Translation and orientation errors were analysed with respect to precision thresholds of 0.25~m, 0.5~m, and 5.0~m for translation, and 2$^\circ$, 5$^\circ$, and 10$^\circ$ for orientation.
To evaluate the vision-based localization module, we used 625 geo-referenced database images to generate a 3D point cloud with the COLMAP framework. A separate sequence of 125 query images, captured on a different day, was used for testing. 
The localization accuracy was assessed using translation and orientation errors, following established 6-DoF outdoor localization benchmarks ~\cite{sattler2018benchmarking}. Translation and orientation errors were analysed with respect to precision thresholds of 0.25~m, 0.5~m, and 5.0~m for translation, and 2$^\circ$, 5$^\circ$, and 10$^\circ$ for orientation. The system localized 76.0\% of the queries within 5.0~m translation error and 100.0\% within 5$^\circ$ orientation error. No queries were localized within the strict translation thresholds of 0.25~m and 0.5~m. These results reflect the inherent challenges of aerial image-based localization in large-scale mining environments.

The Cumulative Distribution Functions (CDFs) of the translation and orientation errors are presented in Fig.~\ref{fig:cdf_translation} and Fig.~\ref{fig:cdf_orientation}, respectively. The orientation CDF exhibits a sharp rise, indicating reliable heading estimation across queries. In contrast, translation errors exhibit a broader distribution. In addition, summary statistics of localization errors are provided in Table~\ref{tab:error_summary}. The mean translation error was measured at 3.69~m with a standard deviation of 1.79~m, while the mean orientation error was 1.77$^\circ$ with a standard deviation of 0.77$^\circ$. These results demonstrate the system’s suitability for UAV-based inspection and mapping tasks in GNSS-denied environments, while highlighting areas for potential improvement in translation precision.

%\subsection{System’s Performance}
%A visual localization module was deployed in GNSS-denied scenarios to validate the system's positioning accuracy. The approach combines visual place recognition and geometric pose estimation (PnP), anchored on a georeferenced COLMAP model. In cases where direct registration failed, a fallback SfM method was employed to preserve continuity.

%Across the test dataset, the system achieved an average translation error of 3.69 meters and an orientation error of 1.77 degrees. These values indicate a high level of spatial precision suitable for UAV-based inspection and mapping tasks in industrial environments. The distribution of localisation errors is presented in Figure \ref{fig:PLOT} and summarised in Table~\ref{tab:loc_errors}.
%\begin{table}[ht]
%\centering
%\caption{Localization Error Summary}
%\begin{tabular}{lcc}
%\hline
%\textbf{Metric} & \textbf{Mean} & \textbf{Std Dev} \\
%\hline
%Translation Error (m) & 3.69 & 1.79 \\
%Orientation Error (°) & 1.77 & 0.77 \\
%\hline
%\end{tabular}
%\label{tab:loc_errors}
%\end{table}

%\begin{figure}[htbp]
%\centerline{ \includegraphics[width=0.5\textwidth]{localization_error_boxplot.png}}
%\caption{Boxplot showing the distribution of localization errors across UAV test sequences}
%\label{fig:PLOT}
%\end{figure}
\subsection{Limitations of the Experimental Deployment}

The proposed system was evaluated in a single real-world scenario: the TERNA MAG Open-pit Mining site. While this site provided a representative environment for aerial mapping and object detection under typical industrial conditions, the generalizability of the approach remains to be validated across more diverse operational settings.

Specifically, the test environment featured relatively stable lighting, moderate terrain elevation, and favorable UAV flight conditions. As such, the system performance was not assessed in environments with challenging characteristics, such as low-light underground tunnels, high dust density, GNSS-shadowed areas, or irregular terrain.
Additionally, certain operational constraints affected data collection. These included:
\begin{itemize}
    \item Restricted access to specific site regions for safety reasons.
    \item Limited UAV flight altitude due to regulatory compliance.
    \item Time-of-day constraints affecting illumination conditions.
\end{itemize}

Although the obtained results are promising, further validation in heterogeneous mining contexts is necessary to evaluate the robustness, adaptability, and potential edge-case failures of the proposed framework. This will constitute a key focus of future work.

\section{Discussion}
The results of this study highlight the potential of combining visual navigation, object detection, and geo-referenced 3D reconstruction to address the challenges of localization and mapping in mining environments, particularly in GNSS-denied areas. The use of COLMAP for geo-referenced modelling, along with the application of PnP for pose estimation, proved to be an effective strategy. Additionally, the fallback mechanism through SfM reconstruction ensured trajectory continuity, even when certain frames could not be registered due to occlusion or insufficient features.

Despite the promising functionality of the developed components, several site-specific limitations were identified. The collected imagery contains a limited number of object classes, restricting dataset diversity and affecting model robustness. The \textit{human} class is notably underrepresented, and due to high-altitude UAV flights, humans appear too small for reliable detection. Moreover, the site’s homogeneous terrain, dominated by soil and dust, introduces visual ambiguities that make it difficult to distinguish among similar classes, such as \textit{trucks} transporting soil.

A key strength of the proposed system is its seamless integration into an interactive Metaverse environment, which not only enables real-time visualization but also enhances situational awareness and decision-making. However, certain aspects, such as localisation robustness in dynamic scenes, near real-time performance, and compatibility with proprietary UAV systems, require further development. The current prototype provides a solid foundation for future scale-up efforts, with optimisation strategies focusing on modular near real-time integration, computational efficiency, and improved robustness in visually degraded or complex real-world mining conditions. Additionally, future work should explore the incorporation of complementary sensor modalities, such as ground-based IoT devices, to enrich spatial context and enhance system adaptability across different mining scenarios. Furthermore, the vision-based localization pipeline remains susceptible to environmental challenges, such as dust, occlusion, and low-light conditions, commonly encountered in open-pit and underground mines. Additionally, the average translation error of 3.69 m, although acceptable for mapping, may not suffice for precision-critical tasks such as autonomous navigation or collision avoidance. The computational demands of SfM and YOLOv8l, particularly when operating on large datasets or in near real-time, also pose deployment challenges in edge-constrained environments.

Based on the above, our plans for further improving the proposed system include: i) exploring transformer-based detection models to improve robustness under variable visual conditions, ii) incorporating complementary sensing modalities, such as thermal or hyperspectral imaging and ground-level IoT data, and iii) optimizing the pipeline for real-time operation using embedded AI platforms like NVIDIA Jetson. Furthermore, expanding the validation to diverse mining environments with varying terrain and lighting characteristics will help assess the generalizability and scalability of the framework.
\section{Conclusion}
In conclusion, the proposed framework presents a comprehensive and adaptable solution for UAV-based navigation and object localisation in complex and often GNSS-challenged mining environments. Using vision-based positioning, object detection from deep learning and 3D geospatial modeling, the system enables high-precision mapping and real-time situational awareness. Its layered design ensures robust localization even in difficult conditions, using visual place recognition and SfM techniques to compensate for the absence or unreliability of GNSS signal. The integration of this approach into an Industrial Metaverse ecosystem further enhances its practical value, allowing for interactive monitoring, predictive analytics, and informed decision-making in virtual mining spaces. However, several challenges remain to be addressed. The system's localization accuracy, though effective for general mapping, may not yet support safety-critical tasks requiring sub-meter precision. Environmental factors such as low lighting, dust, and occlusions can impair the visual pipeline, while the computational demands of deep models and 3D reconstruction limit real-time applicability in edge-constrained deployments. Future directions include incorporating transformer-based vision models and multimodal sensors (e.g., thermal or hyperspectral cameras, and ground-based IoT) to improve robustness and adaptability. Optimizing the full pipeline for execution on embedded platforms will enable real-time performance in the field. Finally, expanding evaluation to diverse mining sites with varied environmental conditions will help validate the generalizability and industrial readiness of the proposed framework.

\section*{ACKNOWLEDGEMENT}This research has been co-financed by the European Health and Digital Executive Agency (HADEA), under the powers delegated by the European Commission (project code: MASTERMINE-101091895).

\bibliography{references}

\begin{thebibliography}{10}
\providecommand{\url}[1]{#1}
\csname url@samestyle\endcsname
\providecommand{\newblock}{\relax}
\providecommand{\bibinfo}[2]{#2}
\providecommand{\BIBentrySTDinterwordspacing}{\spaceskip=0pt\relax}
\providecommand{\BIBentryALTinterwordstretchfactor}{4}
\providecommand{\BIBentryALTinterwordspacing}{\spaceskip=\fontdimen2\font plus
\BIBentryALTinterwordstretchfactor\fontdimen3\font minus \fontdimen4\font\relax}
\providecommand{\BIBforeignlanguage}[2]{{%
\expandafter\ifx\csname l@#1\endcsname\relax
\typeout{** WARNING: IEEEtran.bst: No hyphenation pattern has been}%
\typeout{** loaded for the language `#1'. Using the pattern for}%
\typeout{** the default language instead.}%
\else
\language=\csname l@#1\endcsname
\fi
#2}}
\providecommand{\BIBdecl}{\relax}
\BIBdecl

\bibitem{stothard2023application}
P.~Stothard and R.~Shirani~Faradonbeh, ``Application of uavs in the mining industry and towards an integrated uav-ai-mr technology for mine rehabilitation surveillance,'' \emph{Mining Technology}, vol. 132, no.~2, pp. 65--88, 2023.

\bibitem{flores2024technological}
R.~O. Flores-Casta{\~n}eda, S.~Olaya-Cotera, M.~L{\'o}pez-Porras, E.~Tarme{\~n}o-Juscamaita, and O.~Iparraguirre-Villanueva, ``Technological advances and trends in the mining industry: a systematic review,'' \emph{Mineral Economics}, pp. 1--16, 2024.

\bibitem{ghahramanieisalou2024digital}
M.~Ghahramanieisalou and J.~Sattarvand, ``Digital twins and the mining industry,'' 2024.

\bibitem{thangiartificial}
D.~S. Thangi, ``Artificial intelligence models for remote sensing applications,'' in \emph{Artificial Intelligence Techniques for Sustainable Development}.\hskip 1em plus 0.5em minus 0.4em\relax CRC Press, pp. 201--217.

\bibitem{10355694}
I.~T. Papapetros, K.~M. Oikonomou, I.~Kansizoglou, K.~A. Tsintotas, and A.~Gasteratos, ``Semantic-based visual vocabulary for loop closure detection,'' in \emph{2023 IEEE International Conference on Imaging Systems and Techniques (IST)}, 2023, pp. 1--5.

\bibitem{10092938}
K.~M. Oikonomou, I.~Kansizoglou, and A.~Gasteratos, ``A hybrid reinforcement learning approach with a spiking actor network for efficient robotic arm target reaching,'' \emph{IEEE Robotics and Automation Letters}, vol.~8, no.~5, pp. 3007--3014, 2023.

\bibitem{zhang2022parallel}
H.~Zhang, G.~Luo, Y.~Li, and F.-Y. Wang, ``Parallel vision for intelligent transportation systems in metaverse: Challenges, solutions, and potential applications,'' \emph{IEEE Transactions on Systems, Man, and Cybernetics: Systems}, vol.~53, no.~6, pp. 3400--3413, 2022.

\bibitem{bansal2025exploring}
D.~Bansal, N.~Bhattacharya, and P.~Shandilya, ``Exploring applications: Intelligent drones and robots in industrial settings,'' in \emph{Building Embodied AI Systems: The Agents, the Architecture Principles, Challenges, and Application Domains}.\hskip 1em plus 0.5em minus 0.4em\relax Springer, 2025, pp. 159--180.

\bibitem{nalmpant2025framework}
D.~M. Nalmpant-Sarikaki, A.~I. Theocharis, N.~C. Koukouzas, and I.~E. Zevgolis, ``A framework for effective multi-hazard risk assessment in post-mining areas,'' \emph{Safety}, vol.~11, no.~1, p.~18, 2025.

\bibitem{du2025industrial}
H.~Du, L.~Chan, J.~Tong, R.~Raad, F.~Naghdy, Q.~Guo, Y.~Yu, M.~R. Islam, F.~Tubbal, M.~Ros \emph{et~al.}, ``Industrial progress of robotic automation in mining applications: A survey,'' \emph{Mining, Metallurgy \& Exploration}, pp. 1--20, 2025.

\bibitem{asadzadeh2022uav}
S.~Asadzadeh, W.~J. de~Oliveira, and C.~R. de~Souza~Filho, ``Uav-based remote sensing for the petroleum industry and environmental monitoring: State-of-the-art and perspectives,'' \emph{Journal of Petroleum Science and Engineering}, vol. 208, p. 109633, 2022.

\bibitem{10.1007/978-3-030-34995-0_17}
C.~Sevastopoulos, K.~M. Oikonomou, and S.~Konstantopoulos, ``Improving traversability estimation through autonomous robot experimentation,'' in \emph{Computer Vision Systems}, D.~Tzovaras, D.~Giakoumis, M.~Vincze, and A.~Argyros, Eds.\hskip 1em plus 0.5em minus 0.4em\relax Cham: Springer International Publishing, 2019, pp. 175--184.

\bibitem{said2021application}
K.~O. Said, M.~Onifade, J.~M. Githiria, J.~Abdulsalam, M.~O. Bodunrin, B.~Genc, O.~Johnson, and J.~M. Akande, ``On the application of drones: a progress report in mining operations,'' \emph{International Journal of Mining, Reclamation and Environment}, vol.~35, no.~4, pp. 235--267, 2021.

\bibitem{sarantinoudis2024applications}
N.~Sarantinoudis, N.~Vitzilaios, and G.~Arampatzis, ``Applications of digital twins in uavs,'' in \emph{2024 International Conference on Unmanned Aircraft Systems (ICUAS)}.\hskip 1em plus 0.5em minus 0.4em\relax IEEE, 2024, pp. 450--457.

\bibitem{elbazi2023digital}
N.~Elbazi, H.~El~Hadraoui, O.~Laayati, A.~El~Maghraoui, A.~Chebak, and M.~Mabrouki, ``Digital twin in mining industry: a study on automation commissioning efficiency and safety implementation of a stacker machine in an open-pit mine,'' in \emph{2023 5th Global Power, Energy and Communication Conference (GPECOM)}.\hskip 1em plus 0.5em minus 0.4em\relax IEEE, 2023, pp. 548--553.

\bibitem{uddin2023landing}
J.~Uddin, M.~F. Wadud, R.~Ashrafi, M.~G.~R. Alam, and M.~K. Rhaman, ``Landing with confidence: the role of digital twin in uav precision landing,'' in \emph{2023 10th International Conference on Recent Advances in Air and Space Technologies (RAST)}.\hskip 1em plus 0.5em minus 0.4em\relax IEEE, 2023, pp. 1--6.

\bibitem{papapetros2022visual}
I.~T. Papapetros, V.~Balaska, and A.~Gasteratos, ``Visual loop-closure detection via prominent feature tracking,'' \emph{Journal of Intelligent \& Robotic Systems}, vol. 104, no.~3, p.~54, 2022.

\bibitem{cranford2023conceptual}
R.~Cranford, ``Conceptual application of digital twins to meet esg targets in the mining industry,'' \emph{Frontiers in Industrial Engineering}, vol.~1, p. 1223989, 2023.

\bibitem{zhang2024rotator}
Y.~Zhang, S.~Du, and H.~He, ``Rotator-yolov5: Improved yolov5 for vehicle and vessel detection in uav images,'' in \emph{2024 Fourth International Conference on Digital Data Processing (DDP)}.\hskip 1em plus 0.5em minus 0.4em\relax IEEE, 2024, pp. 156--161.

\bibitem{elbazi2023digitalb}
N.~Elbazi, A.~Tigami, O.~Laayati, A.~El~Maghraoui, A.~Chebak, and M.~Mabrouki, ``Digital twin-enabled monitoring of mining haul trucks with expert system integration: A case study in an experimental open-pit mine,'' in \emph{2023 5th Global Power, Energy and Communication Conference (GPECOM)}.\hskip 1em plus 0.5em minus 0.4em\relax IEEE, 2023, pp. 168--174.

\bibitem{singh2022applications}
M.~Singh, R.~Srivastava, E.~Fuenmayor, V.~Kuts, Y.~Qiao, N.~Murray, and D.~Devine, ``Applications of digital twin across industries: A review,'' \emph{Applied Sciences}, vol.~12, no.~11, p. 5727, 2022.

\bibitem{mitroudas2024light}
T.~Mitroudas, V.~Balaska, A.~Psomoulis, and A.~Gasteratos, ``Light-weight approach for safe landing in populated areas,'' in \emph{2024 IEEE International Conference on Robotics and Automation (ICRA)}.\hskip 1em plus 0.5em minus 0.4em\relax IEEE, 2024, pp. 10\,027--10\,032.

\bibitem{10355707}
\BIBentryALTinterwordspacing
------, ``Multi-criteria decision making for autonomous {UAV} landing,'' in \emph{{IEEE} International Conference on Imaging Systems and Techniques, {IST} 2023, Copenhagen, Denmark, October 17-19, 2023}.\hskip 1em plus 0.5em minus 0.4em\relax {IEEE}, 2023, pp. 1--5. [Online]. Available: \url{https://doi.org/10.1109/IST59124.2023.10355707}
\BIBentrySTDinterwordspacing

\bibitem{balaska2021enhancing}
V.~Balaska, L.~Bampis, I.~Kansizoglou, and A.~Gasteratos, ``Enhancing satellite semantic maps with ground-level imagery,'' \emph{Robotics and Autonomous Systems}, vol. 139, p. 103760, 2021.

\bibitem{nguyen2022application}
L.~Q. Nguyen, M.~T. Dang, L.~K. Bui, Q.~B. Ngoc, and T.~X. Tran, ``Application of unmanned aerial vehicles for surveying and mapping in mines: a review,'' in \emph{International Conference on Geo-Spatial Technologies and Earth Resources}.\hskip 1em plus 0.5em minus 0.4em\relax Springer, 2022, pp. 1--22.

\bibitem{li2024transformer}
X.~Li, H.~Ding, H.~Yuan, W.~Zhang, J.~Pang, G.~Cheng, K.~Chen, Z.~Liu, and C.~C. Loy, ``Transformer-based visual segmentation: A survey,'' \emph{IEEE transactions on pattern analysis and machine intelligence}, 2024.

\bibitem{he2025diffusion}
C.~He, Y.~Shen, C.~Fang, F.~Xiao, L.~Tang, Y.~Zhang, W.~Zuo, Z.~Guo, and X.~Li, ``Diffusion models in low-level vision: A survey,'' \emph{IEEE Transactions on Pattern Analysis and Machine Intelligence}, 2025.

\bibitem{doi:10.1080/25726668.2020.1786298}
H.~Jang and E.~Topal, ``Transformation of the australian mining industry and future prospects,'' \emph{Mining Technology}, vol. 129, no.~3, pp. 120--134, 2020.

\bibitem{Azhari2017A}
F.~Azhari, S.~Kiely, C.~Sennersten, C.~Lindley, M.~Matuszak, and S.~Hogwood, ``A comparison of sensors for underground void mapping by unmanned aerial vehicles,'' pp. 419--430, 2017.

\bibitem{jozkow2021monitoring}
G.~J{\'o}{\'z}k{\'o}w, A.~Walicka, and A.~Borkowski, ``Monitoring terrain deformations caused by underground mining using uav data,'' \emph{The International Archives of the Photogrammetry, Remote Sensing and Spatial Information Sciences}, vol.~43, pp. 737--744, 2021.

\bibitem{to2021drone}
A.~To, M.~Liu, M.~Hazeeq Bin Muhammad~Hairul, J.~G. Davis, J.~S. Lee, H.~Hesse, and H.~D. Nguyen, ``Drone-based ai and 3d reconstruction for digital twin augmentation,'' in \emph{International conference on human-computer interaction}.\hskip 1em plus 0.5em minus 0.4em\relax Springer, 2021, pp. 511--529.

\bibitem{akbulut2025review}
N.~K. Akbulut, A.~Anani, and S.~O. Adewuyi, ``Review of virtual reality integration for safer \& efficient mining operations,'' \emph{IEEE Access}, 2025.

\bibitem{10017382}
K.~Liu, L.~Chen, L.~Li, H.~Ren, and F.-Y. Wang, ``Metamining: Mining in the metaverse,'' \emph{IEEE Transactions on Systems, Man, and Cybernetics: Systems}, vol.~53, no.~6, pp. 3858--3867, 2023.

\bibitem{redmon2016you}
J.~Redmon, S.~Divvala, R.~Girshick, and A.~Farhadi, ``You only look once: Unified, real-time object detection,'' in \emph{Proceedings of the IEEE conference on computer vision and pattern recognition}, 2016, pp. 779--788.

\bibitem{schoenberger2016sfm}
J.~L. Sch\"{o}nberger and J.-M. Frahm, ``Structure-from-motion revisited,'' in \emph{Conference on Computer Vision and Pattern Recognition (CVPR)}, 2016.

\bibitem{lowe2004distinctive}
D.~G. Lowe, ``Distinctive image features from scale-invariant keypoints,'' \emph{International journal of computer vision}, vol.~60, pp. 91--110, 2004.

\bibitem{sattler2018benchmarking}
T.~Sattler, W.~Maddern, C.~Toft, A.~Torii, L.~Hammarstrand, E.~Stenborg, D.~Safari, M.~Okutomi, M.~Pollefeys, J.~Sivic \emph{et~al.}, ``Benchmarking 6dof outdoor visual localization in changing conditions,'' in \emph{Proceedings of the IEEE conference on computer vision and pattern recognition}, 2018, pp. 8601--8610.

\end{thebibliography}
\bibliographystyle{IEEEtran}

\begin{IEEEbiography}[{\includegraphics[width=1in,height=1.25in,clip,keepaspectratio]{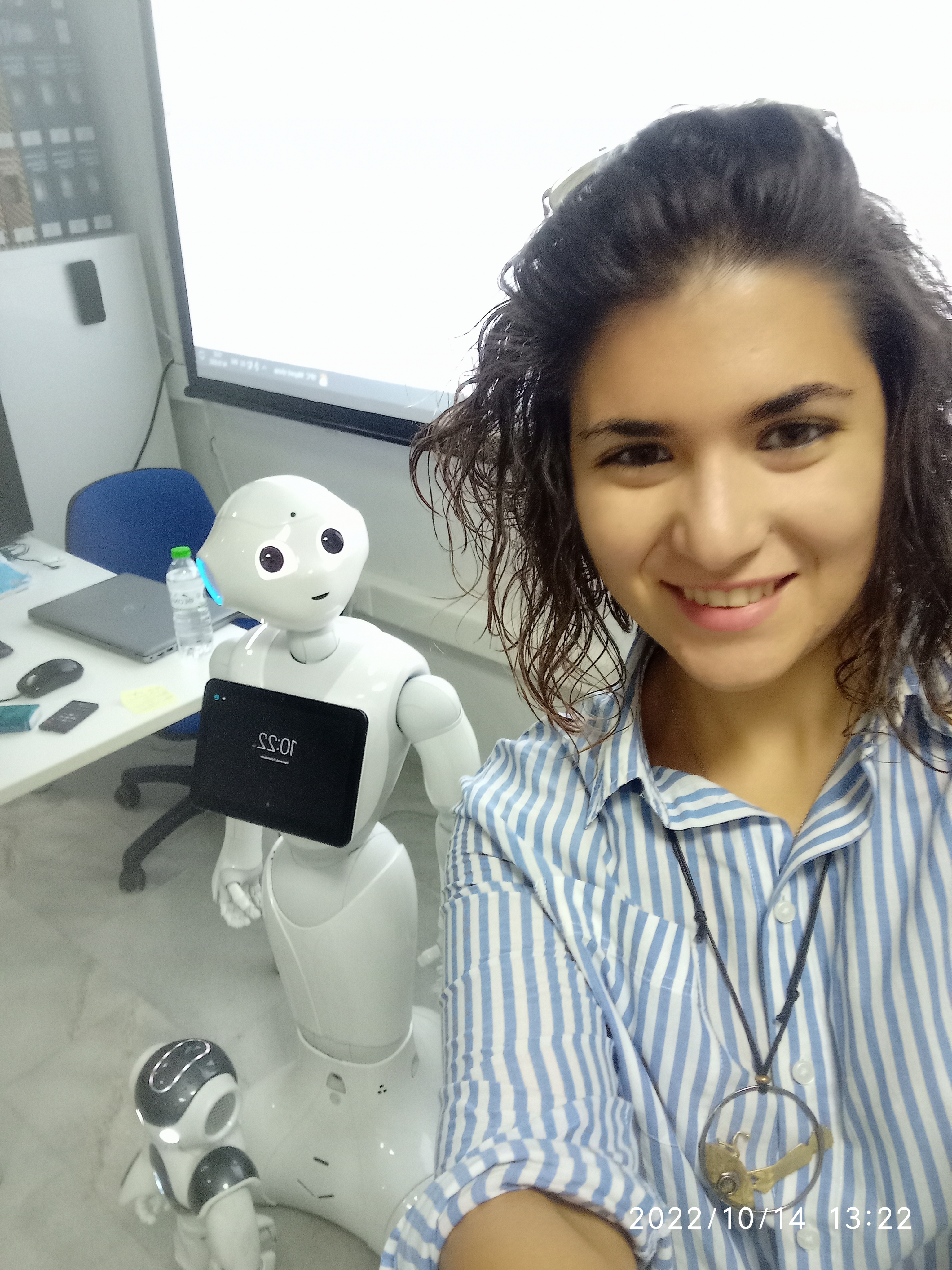}}]{Dr. Balaska Vasiliki} is a Postdoctoral Researcher at the Democritus University of Thrace, Xanthi, Greece, working with Prof. Antonios Gasteratos, and also is a Researcher on the Control Systems and Robotics Laboratory of the Hellenic Mediterranean University. She received her PhD in developing semantic mapping methods and their roles in robotics(2021). 
She has been awarded with the IKY fellowship (2018) for her postgraduate studies. She holds a Diploma in Production and Management Engineering (2017) from the Department of Production and Management Engineering of the DUTH.
Her research interests span the themes of semantic mapping, navigation, and self-localization. 
Also, her publications have appeared in~\emph{IEEE Transactions on Instrumentation and Measurement, Journal of Engineering Applications of Artificial Intelligence, Journal of Intelligent and Robotic Systems, Journal of Robotics and Autonomous Systems}

\end{IEEEbiography}
\begin{IEEEbiography}
[{\includegraphics[width=1in,height=1.25in,clip,keepaspectratio]{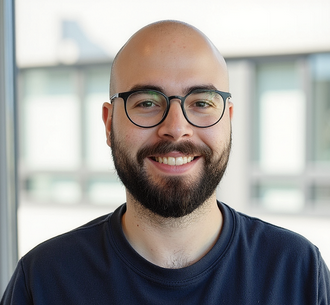}}]{ Ioannis Tsampikos Papapetros} 
received his diploma from the Department of Production and Management Engineering at the Democritus University of Thrace (DUTh), Xanthi, Greece, in 2019. He is a Ph.D. candidate at the Laboratory of Robotics and Automation, working in the field of visual place recognition and robot localization. His broader research interests lie in robotic perception and autonomous navigation.
\end{IEEEbiography}

\begin{IEEEbiography}[{\includegraphics[width=1in,height=1.25in,clip,keepaspectratio]{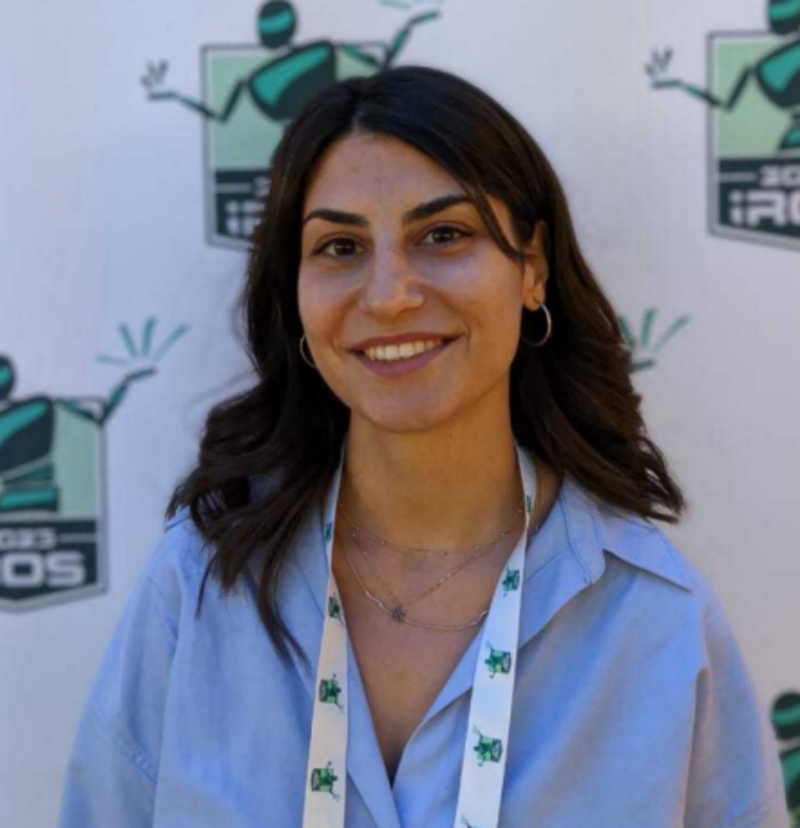}}]{ Katerina Maria Oikonomou} is a Ph.D. Candidate at the Democritus University of Thrace, Xanthi, Greece, working with Prof. Antonios Gasteratos. She received her Masters’s Degree in Robotics and Automation Systems from the National Technical University of Athens, Greece, having her Master’s thesis achieved at the University of Bremen, Germany. She also holds a Bachelor's Degree in Physics from the University of Patras, Greece, having followed the informatics, electronics, and signal processing specialization. Her research interests include robotics, human-robot interaction, and spiking neural networks. More details about her are available at: 
https://robotics.pme.duth.gr/koikonomou/
\end{IEEEbiography}
\begin{IEEEbiography}[{\includegraphics[width=1in,height=1.25in, clip, keepaspectratio]{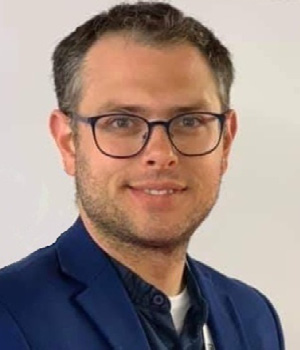}}]{Asst. Prof. Loukas Bampis} received his Diploma from the Department of Electrical and Computer Engineering in 2013 and his Ph.D. from the Department of Production and Management Engineering, both from the Democritus University of Thrace (DUTH). He is currently an Assistant Professor at the Department of Electrical and Computer Engineering, DUTh, and his research interests primarily focus on designing and integrating robotics and mechatronics systems within modern industries, with a specific emphasis on deep learning techniques and sensor modeling for autonomous mapping and localization applications. He has been involved in several research projects funded by the Greek Government and the European Union. He is the author of more than 50 publications in international scientific journals and conferences, for which he has received over 1000 citations. He is a member of IEEE and serves as a reviewer for reputable international scientific journals and conferences. More details about him are available at:https://utopia.duth.gr/lbampis/. 
\end{IEEEbiography}
\begin{IEEEbiography}[{\includegraphics[width=1in,height=1.25in, clip, keepaspectratio]{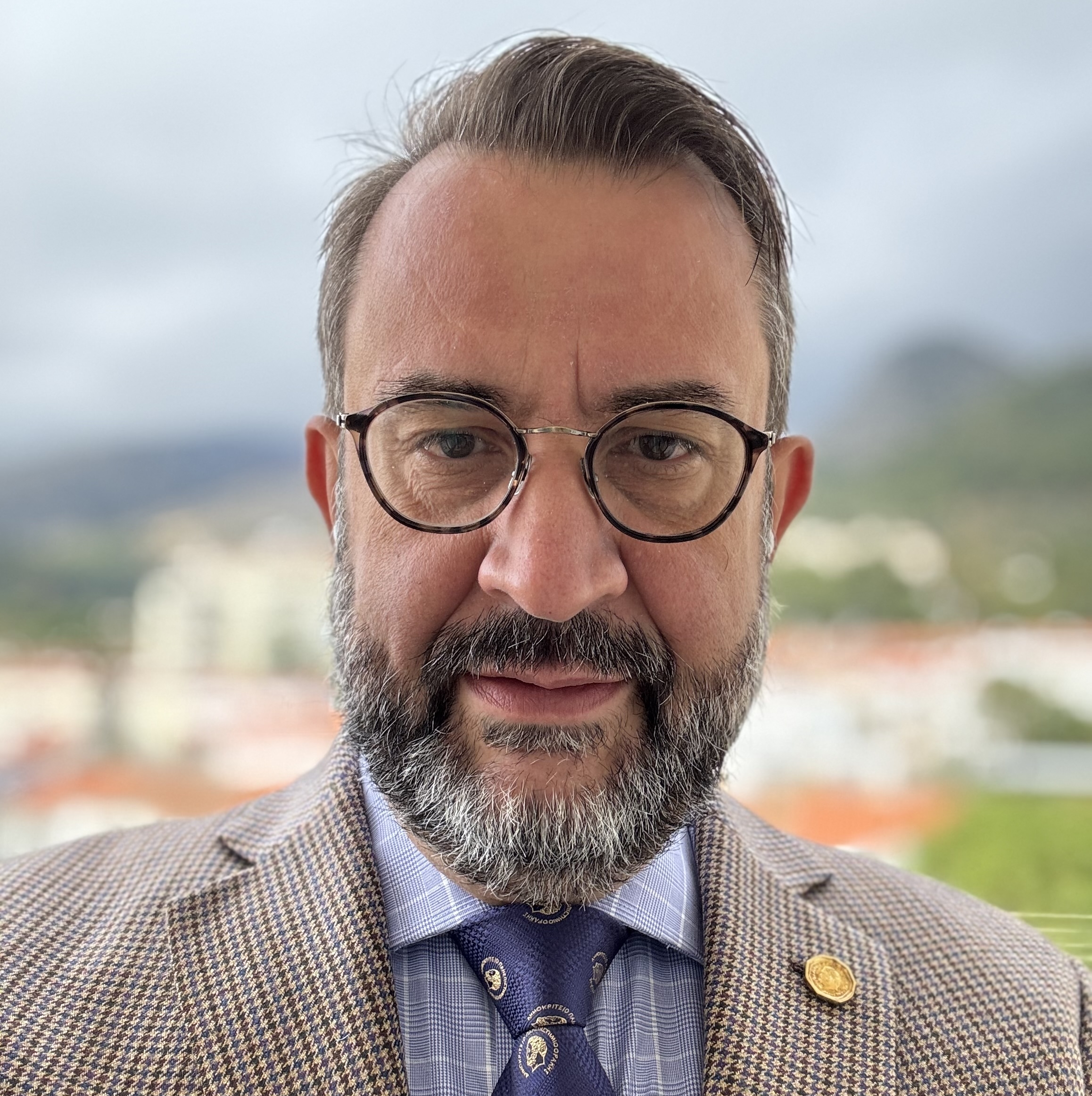}}]{Prof. Antonios Gasteratos} (Senior Member, IEEE) received the M.Eng. and Ph.D. degrees from the Department of Electrical and Computer Engineering, Democritus University of Thrace (DUTh), Xanthi, Greece, in 1994 and 1998, respectively.
From 1999 to 2000, he was a Visiting Researcher with the Laboratory of Integrated Advanced Robotics (LIRALab), DIST, University of Genoa, Genoa, Italy. He is currently a Professor at the Department of Production and Management Engineering at DUTH.
He is also the Director of the Laboratory of Robotics and Automation at DUTH and teaches courses in robotics, automatic control systems, electronics, mechatronics, and computer vision. He has authored over 220 papers in books, journals, and conferences.
His research interests include mechatronics and robot vision. Dr. Gasteratos is a Fellow member of IET.
He has served as a reviewer for numerous scientific journals and international conferences.
He is a Subject Editor of Electronics Letters and an Associate Editor of the International Journal of Optomecatronics.
He has organized/co-organized several international conferences. More details about him are available at:
http://robotics.pme.duth.gr/antonis.
\end{IEEEbiography}

\vfill

\end{document}